%% file: gluonic_skin_prd.tex
\documentclass[aps,prd,reprint,superscriptaddress,amsmath,amssymb,nofootinbib,floatfix]{revtex4-2}

\usepackage{graphicx}
\usepackage{xcolor}
\usepackage{bm}
\usepackage{booktabs}
\usepackage{placeins}
\usepackage[colorlinks=true,linkcolor=blue,citecolor=blue,urlcolor=blue]{hyperref}

\newcommand{\Jpsi}{J/\psi}
\newcommand{\Ups}{\Upsilon}
\newcommand{\drnp}{\Delta r_{np}}
\newcommand{\bsq}{\langle b_\perp^2\rangle}
\newcommand{\rsq}{\langle r^2\rangle}
\newcommand{\kappag}{\kappa}

\newcommand{\dd}{\mathrm{d}}
\newcommand{\Pb}{^{208}\mathrm{Pb}}
\newcommand{\Ca}{^{48}\mathrm{Ca}}
\newcommand{\Sn}{^{124}\mathrm{Sn}}

\input{data/numbers}

\begin{document}
\raggedbottom

\title{Neutron-Skin Sensitivity and Calibration Requirements in Coherent Quarkonium Diffraction}

\author{Lei Wang}
\email{leiwang@ccnu.edu.cn}
\affiliation{Key Laboratory of Quark and Lepton Physics (MOE) and Institute of Particle Physics, Central China Normal University, Wuhan, Hubei 430079, China}

\date{October 27, 2026}

\begin{abstract}
Coherent quarkonium spectra probe the spatial gluon distribution and, under specified assumptions, the neutron skin of $\Pb$.
For three accessible channels at the Electron--Ion Collider, a joint spectral fit with specified selected-event totals and auxiliary constraints gives a coherent-only local uncertainty of $\CoherentSigma$ fm, increasing to $\BackgroundSigma$ fm with a residual incoherent yield equal to 5\% of the coherent yield.
To interpret the spectra in terms of a neutron radius, we examine separately the finite-dipole response and the relation between gluon and nucleon distributions.
We test response changes and calibration offsets in the presence of background using direct nonlinear Poisson fits.
The full spectrum distinguishes a specified density-dependent gluon deformation, whereas a constructed displacement of the gluon edge is exactly degenerate with the neutron skin.
An independent 0.03 fm calibration uncertainty in this direction raises the background-inclusive local width to $\EdgeThreeSigma$ fm; the construction does not predict the physical size of such a displacement.
The results show how the conditional sensitivity depends on an external calibration of the spatial gluon distribution.
\end{abstract}

\maketitle

\section{Introduction}

The neutron skin of a nucleus is a compact measure of how isovector nuclear forces distribute neutrons relative to protons.
In energy-density-functional (EDF) analyses it correlates with the density slope of the symmetry energy and therefore with the pressure of neutron-rich matter near saturation~\cite{Brown:2000pd,Typel:2001sy,Horowitz:2001ya,Reinhard:2010zz,RocaMaza:2011pm,Tsang:2012se,Thiel:2019tkm,Reed:2021nqk}.
The same pressure scale enters neutron-star equation-of-state inference through radii and tidal deformabilities~\cite{Lattimer:2000nx,Oertel:2016bki,Riley:2021pdl,Miller:2021qha,LIGOScientific:2017vwq,LIGOScientific:2018cki}.
The experimental situation is, however, unsettled: PREX-II reported a large neutron skin in $\Pb$, whereas CREX found a comparatively thin skin in $\Ca$~\cite{PREX:2021umo,CREX:2022kgg}.
This tension motivates probes with microscopic systematics distinct from parity-violating electron scattering and hadronic reactions.

The Electron--Ion Collider (EIC) offers such a probe through coherent exclusive vector-meson production, one of the central imaging channels identified in the EIC science program~\cite{Accardi:2012qut,AbdulKhalek:2021gbh}.
At high energy, coherent $\Jpsi$ production is sensitive to the Fourier transform of the average small-$x$ gluon thickness of the nucleus~\cite{Ryskin:1992ui,Brodsky:1994kf,Kowalski:2003hm,Kowalski:2006hc}.
The corresponding finite-$x$ process radius is related to the spatial information encoded in gluon generalized parton distributions and, in the short-distance and $x$-integrated limit, to gluon gravitational form factors (GFFs)~\cite{Ji:1996ek,Radyushkin:1997ki}.

The weighted gluon-radius relation and its application to nuclear neutron distributions were proposed by Hagiwara, Tong, and Xiao~\cite{Hagiwara:2024wqz}.
Li \textit{et al.} subsequently studied neutron-skin sensitivity in coherent and incoherent $\Jpsi$ photoproduction in ultraperipheral collisions~\cite{Li:2026neutronskin}.
Building on these proposals, we ask how much neutron-radius information the diffractive pattern retains when the interaction response and the spatial gluon distribution are allowed to vary.
Recent global proton--lead fits and posterior-based nuclear-structure studies provide relevant calibration frameworks~\cite{Mantysaari:2026global,Mantysaari:2026nuclear}.
Here we use a controlled selected-exposure calculation to quantify the sensitivity to those calibrations.

We separate the sensitivity of the spectra from the external calibration needed to interpret it.
The joint fit gives a conditional local skin uncertainty near 0.025 fm for coherent events and 0.031 fm with the specified residual background.
We match the finite-dipole process radius to the nucleon-center radius and test how response calibration affects both precision and bias.
We then compare a density-dependent gluon deformation that the spectral shape can resolve with a constructed edge displacement that it cannot separate from the neutron skin.
Spatially dependent nuclear parton distributions motivate this examination of the leading profile~\cite{Helenius:2012spatial}.
This comparison identifies the spatial information needed from external sources; neither diagnostic deformation is assigned a physical uncertainty.

We use a common numerical model and a joint background likelihood.
Section~\ref{sec:sumrule} states the conditional radius relation; Sec.~\ref{sec:method} defines accessible kinematics, coordinates, and inference inputs.
Sections~\ref{sec:results} and \ref{sec:closure} quantify calibration dependence and joint closure.
The final sections discuss the separate EDF interpretation and the calibration inputs needed for an experimental application.
All event totals are explicit selected-exposure assumptions; luminosity, photon flux, acceptance, and veto efficiencies are not inferred from them.

\section{Radius Sum Rule and Calibrated Skin Extraction}
\label{sec:sumrule}

The gluon momentum gravitational form factor is an energy--momentum-tensor matrix element. In the convention $H^g(x,0,0;\mu)=xg(x;\mu)$, its zero-skewness representation is
\begin{equation}
 A_g(t;\mu)=\int_0^1\dd x\,H^g(x,0,t;\mu).
 \label{eq:strictgff}
\end{equation}
The small-$x$ dipole representation developed in Ref.~\cite{Hagiwara:2024wqz} links its short-distance coefficient to spatial gluon information. It does not make a measurement at one $x$ an evaluation of the full integral in Eq.~(\ref{eq:strictgff}). The observable considered here is instead the transverse process radius
\begin{equation}
 B_V(x,Q^2)=-4\left.\frac{\partial\ln|\mathcal A_V|}{\partial\Delta^2}\right|_{\Delta^2=0},
 \label{eq:processradius}
\end{equation}
with $\Delta$ in inverse length units. The factor of four converts the amplitude slope into a transverse mean-square radius; the cross-section slope with respect to $\Delta^2$ is $B_V/2$, or $B_V/[2(\hbar c)^2]$ with respect to $|t|$ in $\mathrm{GeV}^2$.

The coherent and incoherent Good--Walker moments are~\cite{Good:1960ba,Lappi:2010dd}
\begin{equation}
  \frac{\dd\sigma_{\rm coh}}{\dd t}=\frac{|\langle{\cal A}\rangle|^2}{16\pi},
  \qquad
  \frac{\dd\sigma_{\rm inc}}{\dd t}=
  \frac{\langle |{\cal A}|^2\rangle-|\langle{\cal A}\rangle|^2}{16\pi}.
  \label{eq:goodwalker}
\end{equation}
where the brackets denote an average over nuclear configurations.
The coherent spectrum measures the event-averaged profile, whereas the incoherent spectrum measures its variance.
At leading order in the dipole size the process radius can be matched to a fixed-$x$ additive gluon moment,
\begin{equation}
  B_V(x,Q^2)=B_g^{(0)}(x)+\Delta_{\rm phase}(x,Q^2)
  +\Delta_{\rm sat}(x,Q^2),
  \label{eq:matching}
\end{equation}
where $\Delta_{\rm phase}$ is generated by the finite nonforward dipole phase and $\Delta_{\rm sat}$ by nonlinear scattering.
The strict GFF radius additionally requires the $x$ integration in Eq.~(\ref{eq:strictgff}); no such extrapolation is assumed in the numerical skin analysis.

Under an additive, impulse-like leading-profile assumption, the fixed-$x$ nuclear gluon thickness can be written as a weighted nucleon sum,
\begin{equation}
  T_g(b)=(1-W)t_p(b)+Wt_n(b),\qquad
  W=\frac{\kappag N}{Z+\kappag N},
  \label{eq:thickness}
\end{equation}
where $t_p$ and $t_n$ are unit-normalized proton and neutron center thicknesses, and $\kappag=g_n/g_p$ is the neutron-to-proton gluon charge ratio.
Equivalently, for number-normalized profiles $T_p=Zt_p$ and $T_n=Nt_n$, one may write $T_g=(T_p+\kappag T_n)/(Z+\kappag N)$; this is the convention implemented in the numerical profiles below.
The weighted radius relation and its proposed neutron-skin application follow Ref.~\cite{Hagiwara:2024wqz}. Including a calibrated common intrinsic gluon radius of a bound nucleon gives
\begin{equation}
  B_g^{(0)}=\bsq_{\rm int}+(1-W)\bsq_p+W\bsq_n.
  \label{eq:sumrule}
\end{equation}
Here $\bsq_{\rm int}$ is the common bound-nucleon contribution, while $\bsq_p$ and $\bsq_n$ are the transverse mean-square radii of the proton and neutron center distributions.
The common term is removed only after calibration against the proton gluon radius and the point-proton nuclear radius; it does not cancel algebraically from a single nuclear measurement.
Its calibration uncertainty enters the extracted neutron moment with coefficient $1/W$, but a possible isovector intrinsic difference is not included in the numerical covariance. Such a difference requires a separate bound-nucleon calibration.

For the nearly spherical nuclei considered here, $\bsq=2\rsq/3$.
Defining the matched, intrinsic-subtracted moment $\bar B_g=B_V-\Delta_{\rm phase}-\Delta_{\rm sat}-\bsq_{\rm int}$, the neutron radius follows from
\begin{align}
  \bsq_n&=\bsq_p+\frac{\bar B_g-\bsq_p}{W},
  \label{eq:bninvert}\\
  \drnp&=\sqrt{\frac{3}{2}\bsq_n}-\sqrt{\rsq_p}.
  \label{eq:skinextract}
\end{align}
Equivalently, with $R_g^2=(3/2)\bar B_g$ and $R_p^2=\rsq_p$ denoting the point-proton mean-square radius,
\begin{equation}
  \drnp =
  \left[
    \frac{Z+\kappag N}{\kappag N}R_g^2
    -\frac{Z}{\kappag N}R_p^2
  \right]^{1/2}
  - R_p .
  \label{eq:closedform}
\end{equation}
This closed-form expression is conditional on the matching in Eq.~(\ref{eq:matching}) and on a calibrated common intrinsic radius; it is not a model-independent inversion of the measured spectrum.

The dependence on $\kappag$ is structurally suppressed.
Writing $\Delta B_n=\bsq_n-\bsq_p=(\bar B_g-\bsq_p)/W$ gives
\begin{equation}
  \frac{\delta(\Delta B_n)}{\Delta B_n}
  =-(1-W)\frac{\delta\kappag}{\kappag}.
  \label{eq:kappa}
\end{equation}
For $\Pb$ and $\kappag=1$, $W=N/A\simeq0.61$.
A few-percent prior on $\kappag$ therefore contributes a subleading skin uncertainty.
The principal question is whether the process corrections in Eq.~(\ref{eq:matching}) can be separated from the isovector change of the nuclear edge; isoscalarity of their central values alone is insufficient to establish that separation.

\section{Kinematics and midpoint-consistent amplitude}
\label{sec:method}

We use $E_e=18$ GeV and $E_N=100$ GeV per nucleon as a heavy-ion beam benchmark, giving $s_{eN}\simeq4E_eE_N=7200~\mathrm{GeV}^2$.
This choice is within the heavy-ion energy envelope of the EIC requirements~\cite{EIC:requirements}; it is not a committed run configuration.
The per-nucleon diffractive momentum fraction is
\begin{equation}
 x_{\mathbb P}=\frac{M_V^2+Q^2-t}{W_{\gamma N}^2+Q^2-m_N^2}
 \simeq\frac{M_V^2+Q^2}{y s_{eN}}.
 \label{eq:kinematics}
\end{equation}
We identify the opacity argument $x$ with its forward value and neglect its variation across the fitted $t$ interval, which reaches about 1\% for $\Jpsi$ photoproduction.
Table~\ref{tab:kinematics} gives the resulting $0<y<1$ points, with a channel-dependent $x$ for each meson mass and virtuality.
Photoproduction denotes the quasireal-photon limit; the $Q^2=10~\mathrm{GeV}^2$ entry is a transverse response, conditional on isolating that component. No longitudinal-channel efficiency is assumed.

\begin{table}[!t]
\caption{Forward kinematics and imposed selected coherent yields. The arguments of the channel labels are $Q^2$ in $\mathrm{GeV}^2$; $W$ is in GeV.}
\label{tab:kinematics}
\begin{ruledtabular}\begin{tabular}{lrrrr}
Channel & $x$ & $y$ & $W$ & $N_c$\\
\input{data/kinematics}
\end{tabular}\end{ruledtabular}
\end{table}

Let $\bm r=\bm x_q-\bm x_{\bar q}$ and $\bm b=(\bm x_q+\bm x_{\bar q})/2$.
The longitudinal-momentum-weighted center is
\begin{equation}
 \bm R=z\bm x_q+(1-z)\bm x_{\bar q}
       =\bm b+(z-\tfrac12)\bm r.
\end{equation}
The phase $e^{-i\bm\Delta\cdot\bm R}$ is invariant under the simultaneous exchange $z\to1-z$, $\bm r\to-\bm r$.
For an angle-independent optical amplitude its azimuthal average therefore contains $J_0(qb)J_0[(z-1/2)qr]$, consistent with Refs.~\cite{Hatta:2017phase,Lappi:2020phase}.
Our amplitude, up to a rate factor removed in the shape fit, is
\begin{align}
 \mathcal A_c(q)&\propto e^{-B_{\rm int}q^2/4}
 \int b\dd b\,J_0(qb)\int r\dd r\,\dd z\,\Phi_c(r,z)\nonumber\\
 &\qquad\times J_0[(z-\tfrac12)qr]\,\mathcal N_c(r,b),\qquad
 q=\sqrt{|t|}/(\hbar c),\label{eq:amplitude}\\
 \mathcal N_c&=1-\exp[-\Omega_c(1+\Omega_c/0.7)^\gamma],\label{eq:response}\\
 \Omega_c&=O\frac{\pi^2}{2N_c^{\rm color}}\alpha_s(\mu^2)
 x_cg(x_c,\mu^2)r^2[T_p(b)+\kappag T_n(b)].\label{eq:opacity}
\end{align}
Here $N_c^{\rm color}=3$, $O=e^{\ln O}$, and $\mu^2=\mu_0^2+C(\hbar c)^2/r^2$.
The IP-Sat-motivated input uses LO gluon-only evolution and boosted-Gaussian transverse overlaps~\cite{Kowalski:2003hm,Kowalski:2006hc,Rezaeian:2013ipsat}; it is a specified phenomenological baseline, not a new global IP-Sat fit.
The response parameter $\gamma$ preserves the dilute coefficient and changes the semi-saturated response.
A size nuisance $s_V$ replaces $r$ by $e^{s_V}r$ in the opacity and phase while the integration variable and overlap are held fixed; a shape normalization absorbs the overall Jacobian.

The Woods--Saxon proton and neutron profiles are independently number normalized.
We use $R_p=\sqrt{5.503^2-0.5956}=5.448615$ fm, $\drnp=0.17$ fm, $a_p=0.45$ fm, and $a_n=0.50$ fm.
The half-density radii are solved from these rms radii at each fitted point, rather than keeping the half-density radius fixed when the diffuseness changes.
The charge-to-point-proton conversion follows Ref.~\cite{Horowitz:2012pb}.
The nominal intrinsic moment is $B_{\rm int}=0.370~\mathrm{fm}^2$.
Table~\ref{tab:priors} specifies all auxiliary widths; their central values are the nominal values unless a closure experiment states otherwise.
The common $B_{\rm int}$, $O$, and $\gamma$ across different $x_c$ are calibration hypotheses, not consequences of isospin symmetry.

\begin{table}[!t]
\caption{Gaussian auxiliary standard deviations. The skin and three logarithmic coherent normalizations $\nu_c$ have no Gaussian constraint. The radial-profile tests add parameters explicitly.}
\label{tab:priors}
\begin{ruledtabular}\begin{tabular}{ll}
Parameter & Standard deviation\\
$B_{\rm int}$ & $0.0185~\mathrm{fm}^2$\\
$\ln O,\gamma$ & $0.10,\ 0.15$\\
$a_n,R_p$ & $0.03,\ 0.005$ fm\\
$\kappag,\eta_t$ & $0.05,\ 0.10$\\
$s_{\Jpsi},s_{\Ups}$ & $0.05,\ 0.05$\\
Background $\beta_c,\xi_c$ & $0.20,\ 0.30$\\
Per-bin multiplicative response & $0.005$\\
\end{tabular}\end{ruledtabular}
\end{table}

We integrate a Gaussian resolution kernel of width $0.003e^{\eta_t}~\mathrm{GeV}^2$ over each of 33 bins in $0.002<|t|<0.100~\mathrm{GeV}^2$.
Writing the bin-integrated shape as $S_{ic}$, the coherent mean is
\begin{equation}
 \mu_{ic}=e^{\nu_c}N_c\frac{S_{ic}}{\sum_jS_{jc}}.
 \label{eq:counts}
\end{equation}
All subsequent derivatives include this parameter-dependent normalization.
The local coherent Fisher matrix is $F=D^TV^{-1}D+P$, with $D_{ia}=\partial\mu_i/\partial\theta_a$, $V_{ii}=\mu_i+(0.005\mu_i)^2$, and $P$ the auxiliary precision matrix.
The reported local width is $\sqrt{(F^{-1})_{ss}}$, where $s=\drnp$.
Sparse bins, particularly in $\Ups$, are treated with a Poisson likelihood in the closure tests rather than Gaussian pseudo-counts.

\paragraph*{Use of AI tools.}
OpenAI Codex assisted with code development and debugging, numerical analysis, manuscript restructuring, and drafting.

\section{Spectral sensitivity and calibration requirements}
\label{sec:results}

\subsection{Spectral information and process matching}

\begin{figure}[!t]
\includegraphics[width=\linewidth]{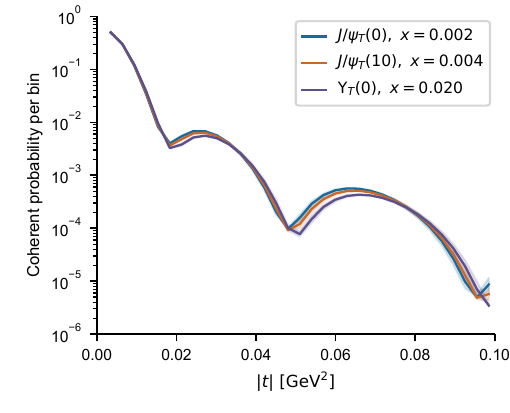}
\caption{Bin-integrated coherent probability distributions for the three accessible channels. Solid curves use $\drnp=0.17$ fm; the lighter envelopes span the illustrative $0.12$--$0.22$ fm variations with all other parameters fixed. The envelopes are parameter variations, not confidence intervals. All curves include the midpoint phase and the same Gaussian resolution.}
\label{fig:spectra}
\end{figure}

The coherent three-channel benchmark retains a local skin sensitivity of $\sigma(\drnp)=\CoherentSigma$ fm after profiling the response and nuclear-geometry parameters with the stated auxiliary constraints.
The skin correlation with $\gamma$ is $\GammaRho$, whereas that with $\ln O$ is $\OpacityRho$.
Thus the response shape is more consequential than its common normalization under the stated auxiliary constraints.
The coherent-only widths are 0.02631 fm for $\Jpsi$ photoproduction alone, 0.02499 fm for the two $\Jpsi$ channels, and 0.02579 fm for $\Jpsi$ photoproduction with $\Ups$. The three-channel width is $\CoherentSigma$ fm.
Figure~\ref{fig:spectra} shows how the spectra depend on the skin.

The leading center moment is $\LeadingMoment~\mathrm{fm}^2$, or $\FullLeadingMoment~\mathrm{fm}^2$ after adding the intrinsic term.
For $\Jpsi_T(0)$ the process moment is $\ProcessMoment~\mathrm{fm}^2$.
For $\Jpsi_T(0)$ the phase and nonlinear corrections are 0.00409 and 0.97792 $\mathrm{fm}^2$, respectively, computed using the same quadrature as the spectra.
The phase term is obtained from the dilute amplitude with the weight $(z-1/2)^2r^2$; the saturation term is the difference between nonlinear and dilute process moments.
These absolute moments are consistent with the nuclear geometry, whose analytic leading relation gives $20.551414~\mathrm{fm}^2$ before the small finite-grid correction.

Tightening $\sigma_\gamma$ to 0.05 gives $\GammaTight$ fm, while loosening it to 0.40 gives $\GammaLoose$ fm.
The auxiliary width 0.15 is illustrative and is not a posterior extracted from proton or nuclear data.
Varying the $\kappag$ auxiliary width from 0.010 to 0.200 changes the coherent width from 0.02435 to 0.02700 fm. Varying the neutron diffuseness width from 0.010 to 0.080 fm changes it from 0.02359 to 0.02489 fm, with the same exposures and phase convention as the main fit.
Replacing the shared $B_{\rm int},\gamma,\ln O$ by three independently calibrated copies, each with the same marginal auxiliary width, gives 0.02136 fm.
This smaller result reflects averaging of independent calibration information; it is not a conservative relaxation of the shared calibration assumption.
The numerical package records the individual replacements and their dependence on auxiliary correlations.

\subsection{Distinguishable profile changes and edge calibration}

The sum rule assumes an additive leading profile. Equation~(\ref{eq:response}) cannot test that assumption because its dilute term remains proportional to $T_A=T_p+\kappag T_n$.
As a restricted density-dependent diagnostic, motivated by the thickness dependence used in spatial nuclear-PDF constructions~\cite{Helenius:2012spatial}, we replace $T_A$ by
\begin{equation}
 T_A^{(\eta)}(b)=\frac{T_A(b)\exp[\eta_A T_A(b)/T_A(0)]}
 {\langle\exp[\eta_A T_A/T_A(0)]\rangle_{T_A}},
 \label{eq:radial}
\end{equation}
where the denominator is a $T_A$-weighted transverse average, preserving $\int\dd^2b\,T_A$.
This ansatz is positive and changes the $r^2$ coefficient; it is not an EPS09s error eigenvector or a fitted shadowing uncertainty.
A common unconstrained $\eta_A$ gives $\RadialFreeSigma$ fm, and allowing an independent $\eta_{A,c}$ in each channel gives $\RadialIndependentSigma$ fm.
The distinct central-density distortion is therefore largely resolved by the full shape in this restricted example.
The omitted-$\eta_A=0.05$ truth is also tested with background in Sec.~\ref{sec:closure}.

\begin{figure*}[!t]
\includegraphics[width=\textwidth]{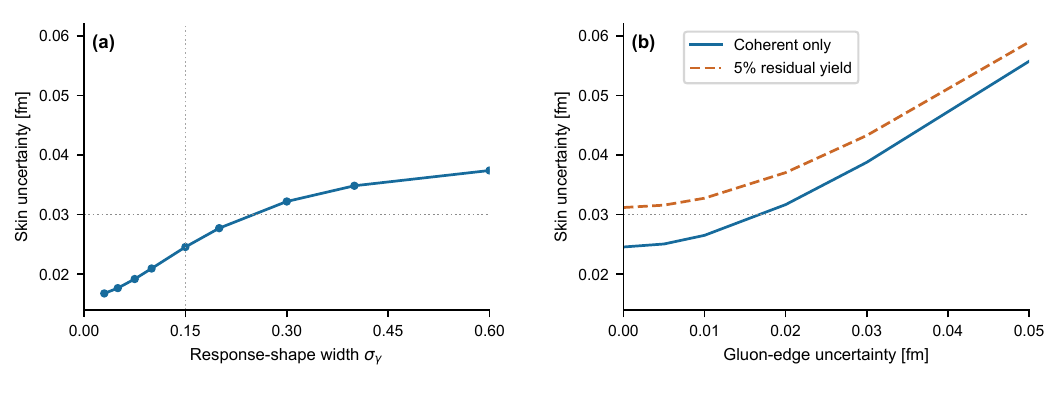}
\caption{Sensitivity to model uncertainties. (a) Conditional coherent skin width versus the response-shape auxiliary width. (b) Width versus the independent gluon-edge uncertainty, with and without a 5\% residual incoherent yield. Curves in (b) follow the exact local degeneracy relation, Eq.~(\ref{eq:edgevariance}); they are not nuclear-PDF uncertainty bands. The horizontal dotted line marks 0.03 fm only as a reference.}
\label{fig:calibration}
\end{figure*}

We next consider a spatial change that requires external calibration even when the full spectrum is available.
Let $s$ denote the nuclear skin and introduce the positive edge modification
\begin{equation}
 R_g^{\rm edge}(b;s,\delta_g)=\frac{T_A(b;s+\delta_g)}{T_A(b;s)},
 \qquad T_A^{\rm edge}=R_g^{\rm edge}T_A.
 \label{eq:edge}
\end{equation}
Both numerator and denominator have the same number normalization, and the proton profile and neutron diffuseness are fixed in this definition.
Every amplitude in the optical calculation then depends on $s+\delta_g$.
The two derivative columns are identical, so the spectra alone cannot separate the nuclear skin from this gluon-edge displacement.
The degeneracy follows by construction from the dependence on $s+\delta_g$.
This construction exhibits a spatial-gluon ambiguity that the spectra leave unresolved. It provides neither a QCD prediction for the modification nor an estimate of its size.
For an independent centered Gaussian calibration of $\delta_g$, the local variance is exactly
\begin{equation}
 \sigma_s^2=\sigma_{s\mid\delta_g=0}^2+\sigma_{\delta_g}^2.
 \label{eq:edgevariance}
\end{equation}
A 0.02 (0.03) fm auxiliary width therefore gives $\EdgeTwoSigma$ ($\EdgeThreeSigma$) fm when the nominal 5\% background is included. Figure~\ref{fig:calibration} shows the dependence on the response-shape and gluon-edge widths.
With no such calibration the skin is unidentifiable along this direction, even with unlimited spectral statistics.

The moment relation expresses the same requirement as
\begin{equation}
 \delta s=\frac{3\,\delta\bar B_g}{4WR_n}
        =\MomentCoefficient~\mathrm{fm}^{-1}\delta\bar B_g.
 \label{eq:tolerance}
\end{equation}
A 0.03 fm tolerance on the induced skin shift corresponds to an unmodeled moment shift of $\MomentTolerance~\mathrm{fm}^2$.
This is a calibration tolerance, not an estimate of nuclear shadowing.
Equations~(\ref{eq:radial}) and (\ref{eq:edge}) illustrate the difference between profile variations constrained by the spectral shape and those requiring external calibration.

\section{Joint background and model-closure experiments}
\label{sec:closure}

\subsection{Counts and auxiliary-data likelihood}

We include the residual incoherent component in the experiments that test response and profile changes.
A Good--Walker fluctuation template $g_i$ is normalized over the fitted bins after resolution smearing.
For each channel we use
\begin{align}
 m_{ic}&=\mu_{ic}+fN_c e^{\beta_c}
 \frac{g_i e^{\xi_c h_i}}{\sum_jg_j e^{\xi_c h_j}},\label{eq:background}\\
 h_i&=(t_i-0.051)/0.049,
\end{align}
with $t_i$ in $\mathrm{GeV}^2$.
Thus $f=0.05$ denotes a residual yield equal to 5\% of the nominal coherent yield, rather than 5\% of the total selected sample.
The background normalizations and shape tilts have independent auxiliary widths 0.20 and 0.30.
The template is a controlled fluctuation model with 4000 configurations, not a detector-specific breakup simulation; its details appear in the Appendix.
The local widths for $f=0,0.02,0.05,0.10,0.20$ are respectively 0.02457, 0.02856, 0.03120, 0.03405, and 0.03809 fm.

Each pseudoexperiment draws $n_{ic}\sim\mathrm{Poisson}(m^{\rm true}_{ic})$ and independent Gaussian auxiliary observations for the constrained physics and background parameters.
For the per-bin 0.5\% response uncertainty we introduce a positive multiplier $b_i$, with a Gaussian auxiliary observation $c_i\sim\mathcal N(1,0.005^2)$ at the generating truth.
The fit minimizes
\begin{align}
 -\ln\mathcal L={}&\sum_i[m_i b_i-n_i\ln(m_i b_i)]
 +\sum_i\frac{(b_i-c_i)^2}{2s_b^2}\nonumber\\
 &+\frac12\sum_a\frac{(\theta_a-a_a)^2}{\sigma_a^2},
 \quad s_b=0.005,\label{eq:likelihood}
\end{align}
up to data-only constants.
For a given physics point the bin multipliers are profiled analytically through
\begin{equation}
 b_i=\frac{c_i-m_i s_b^2+\sqrt{(c_i-m_i s_b^2)^2+4n_i s_b^2}}{2}.
\end{equation}
The remaining 19 parameters are fitted with the direct nonlinear amplitude: the skin, nine constrained physics nuisances, three coherent normalizations, and six background parameters.
The interval is the fitted skin plus or minus one local expected-information standard deviation, evaluated with unit bin-response multipliers and with bin and auxiliary nuisances profiled.
It is a Wald interval with nominal 68.27\% coverage, not an exact Neyman construction.
All quoted coverage results below use 200 independent count-and-auxiliary realizations per truth and direct nonlinear refits.
A local log-response approximation supplies starting points and an additional numerical comparison; it does not supply the reported coverage.

\subsection{Separating calibration offset from model mismatch}

We first vary $\gamma$ to $-0.26$ within the response family of Eq.~(\ref{eq:response}), allowing the fit to adjust the same parameter.
We generate two otherwise identical experiments: one with the $\gamma$ auxiliary measurement centered at $-0.26$, and one centered at zero despite the displaced truth.
The second experiment measures the effect of a systematic auxiliary-calibration offset.
The other auxiliary measurements remain centered on their generating values.

We also consider structural alternatives to the response model.
For the published bCGC truth we integrate the proton amplitude of Ref.~\cite{Rezaeian:2013tka}, using its $m_c=1.4$ GeV parameter row, and embed $\sigma_{dp}(r,x_c)$ through $1-\exp[-T_A\sigma_{dp}/2]$.
The common intrinsic Fourier factor is retained to isolate a response-law change.
The $x=0.020$ use in the $\Ups$ channel extrapolates this small-$x$ parametrization and is a stress construction, not an independently validated prediction at that point.
For the inside-opacity truth, we convolve the nuclear thickness with $G(b)=\exp(-b^2/B_{\rm int})/(\pi B_{\rm int})$ before exponentiation and omit the external intrinsic factor.
This construction is equivalent to external smearing only in the dilute limit.
The two leading-profile truths use Eqs.~(\ref{eq:radial}) and (\ref{eq:edge}) with the added parameter fixed to zero in the fit.
Their generating nuclear skin remains 0.17 fm.

\begin{table*}[!t]
\caption{Joint closure with 5\% residual incoherent yield. Biases and widths are in fm. The Asimov bias is the direct nonlinear fit to the exact mean spectrum and mean auxiliary data. $\langle\sigma\rangle$, mean fitted bias, and coverage are from 200 direct Poisson refits for each row. Coverage errors are one binomial standard error in percentage points, typically about 3 points. The last column is the minimized Asimov deviance including auxiliary pulls; it is not assigned a $p$ value.}
\label{tab:closure}
\begin{ruledtabular}\begin{tabular}{lrrrrr}
Generating truth & Asimov bias & $\langle\sigma\rangle$ & Mean bias & Coverage [\%] & Asimov deviance\\
\input{data/closure_table}
\end{tabular}\end{ruledtabular}
\end{table*}

Table~\ref{tab:closure} distinguishes conditional precision from model error.
The nominal experiment gives \NominalCoverage\% coverage; the correctly centered displaced response gives \GammaCorrectCoverage\%, compared with \GammaOffsetCoverage\% when the auxiliary center is wrong.
The inside-opacity and published-bCGC experiments give \InsideCoverage\% and \BCGCCoverage\%, respectively.
These tests assess specified alternatives with the same leakage likelihood; they do not define a probability distribution over all possible QCD models.
As an additional Asimov-only response diagnostic, replacing the evolved gluon density by its input-scale form gives a bias of $\MVBias$ fm in the joint background fit.
No coverage claim is assigned to that additional case.

The edge-displacement truth produces an Asimov bias of exactly $+0.0300$ fm while retaining an effectively zero Asimov deviance: the wrong nuclear skin reproduces the gluon spectrum exactly.
Its coverage is \EdgeCoverage\% when that displacement is omitted from the fit.
The density-dependent omitted-$\eta_A$ case gives \RadialCoverage\% coverage and behaves differently because it changes a distinguishable shape direction.
A good spectral fit therefore cannot, by itself, validate the additive interpretation of the fitted radius.

\section{Relation to the Symmetry Energy}
\label{sec:symmetry}

Conditional on the process and leading-profile calibrations, the gluonic inference constrains the neutron skin
\begin{equation}
  \drnp=\sqrt{\rsq_n}-\sqrt{\rsq_p},
  \label{eq:skin}
\end{equation}
where $\rsq_n$ and $\rsq_p$ are the neutron and proton mean-square radii.
It does not directly measure the symmetry-energy slope,
\begin{equation}
  L=3n_0\left.\frac{\partial S(n)}{\partial n}\right|_{n_0},
  \label{eq:Ldef}
\end{equation}
but can be mapped to $L$ only through an external nuclear-structure correlation.
The final step therefore requires an external EDF correlation,
\begin{equation}
  \drnp(A,Z)=c_1(A,Z)L+c_0(A,Z),
  \label{eq:Lcorr}
\end{equation}
where $L$ is in MeV and radii are in fm.
Published EDF surveys find strong but model-dependent correlations for $\Pb$ and a weaker relation for $\Ca$~\cite{Reinhard:2010zz,RocaMaza:2011pm,Inakura:2015wca}.
The intercept, slope, and intrinsic scatter must be taken from one common EDF ensemble when EIC, PREX-II, and CREX likelihoods are compared; mixing coefficients inferred from different functional families would create an artificial tension.
We therefore do not quote a combined Pb/Sn/Ca uncertainty on $L$.
In particular, $\Sn$ is retained as a valuable intermediate-mass and isotope-chain target, but no quantitative $\Sn$--$L$ coefficient is used without a dedicated EDF calculation and covariance matrix.

The direct output of the proposed measurement is thus $\drnp$, not $L$.
Its nuclear-structure value lies in providing a third reaction mechanism with different experimental and strong-interaction systematics, so that the compatibility of PREX-II, CREX, and nuclear theory can be tested before a dense-matter interpretation is attempted.
Once a common EDF likelihood is supplied, Eq.~(\ref{eq:Lcorr}) can propagate the measured covariance, including correlations among nuclei and between $c_0$ and $c_1$.

The connection to neutron-star phenomenology is correspondingly indirect.
An EDF interpretation of the skin constrains the symmetry pressure near saturation, $P_{\rm sym}(n_0)\simeq n_0L/3$, not a stellar radius by itself.
The value of the gluonic measurement is that it supplies an independent finite-nucleus constraint in the chain from nuclear radii to dense matter, rather than a direct neutron-star observable.

\section{Discussion}
\label{sec:discussion}

The spectra retain neutron-skin information after the stated response, geometry, and background uncertainties are profiled, giving conditional local widths near 0.025--0.031 fm.
The full shape also distinguishes the specified central-density deformation from a skin change.
These results quantify how coherent diffraction can contribute to the neutron-radius program proposed in Ref.~\cite{Hagiwara:2024wqz} and explored in UPC production in Ref.~\cite{Li:2026neutronskin}.
Some spatial changes are resolved by the spectrum, while others require external information.

Interpreting the spectral information as a neutron radius requires two calibrations.
The finite-dipole response maps a gluon profile into a measured spectrum; the leading-profile calibration relates that gluon profile to neutron and proton center distributions.
The response scans and joint closure tests quantify the sensitivity to the former, including the bias caused by an incorrect auxiliary center.
The constructed edge displacement isolates a direction left unresolved by the spectra. Equation~(\ref{eq:edgevariance}) gives the resulting dependence on the leading-profile calibration.
Its exact degeneracy is a property of the parameterization, not a prediction for the physical nuclear gluon modification.
Establishing the size of such a direction requires external constraints on the spatial gluon distribution and its $x$ dependence.

Proton--nucleus global analyses~\cite{Mantysaari:2026global} and posterior propagation across nuclei~\cite{Mantysaari:2026nuclear} offer routes to replace the illustrative calibration widths with constraints from data.
Applying them requires the channel-to-channel calibration covariance and a common treatment of the nuclear geometry and interaction response.
The EIC and UPC programs have different photon fluxes, breakup selections, and kinematic coverage.
Our selected-exposure benchmark compares these calibration requirements without incorporating those global-fit posteriors.

The two $\Jpsi$ channels alone give 0.02499 fm under the same coherent-only calibration assumptions, so the principal conditional sensitivity is already present without $\Ups$.
The accessible $\Ups$ point serves as a supplementary phenomenological control.
The coherence-length estimate $\ell_c=1/(m_Nx)$ follows from the longitudinal energy mismatch $(M_V^2+Q^2)/(2\nu)$.
With $\hbar c$ restored, the three values are about 105, 53, and 11 fm.
The last is comparable to the Pb diameter, making finite-coherence-length and moderate-$x$ effects relevant for a quantitative use of this channel.
Those effects and their wave-function correlations are outside the frozen-dipole benchmark.

An experimental projection would connect the selected counts to luminosity, photon flux, branching fractions, polarization separation, and efficiency, and replace the Gaussian resolution and controlled background template with channel-specific migration and breakup-veto models~\cite{Chang:2021jpsi,ECCE:2023exclusive}.
The reported widths therefore remain conditional benchmarks, with coverage assessed for the stated count process, auxiliary data, and Wald interval at finite Monte Carlo precision.
An isovector bound-nucleon intrinsic gluon-radius difference, NLO evolution, and $t$-dependent real-part and skewness corrections remain outside the propagated uncertainty.

\section{Conclusions}

Coherent quarkonium spectra at accessible EIC kinematics provide conditional local $\Pb$ neutron-skin sensitivities of $\CoherentSigma$ fm for the specified coherent sample and $\BackgroundSigma$ fm with a 5\% residual incoherent yield.
The spectral shape constrains the specified density-dependent gluon deformation, while the response scans and joint nonlinear Poisson tests quantify the roles of calibration width and calibration bias.
A constructed gluon-edge displacement identifies a complementary requirement: a profile change that mimics the skin must be constrained externally.
For an independent 0.03 fm calibration uncertainty in that direction, the background-inclusive local width is $\EdgeThreeSigma$ fm.
A calibrated neutron-skin inference from these spectra therefore requires external constraints on the spatial gluon distribution.

\paragraph*{Data and code availability.}
No supplemental reproducibility materials are provided with this article.

\begin{acknowledgments}
The author acknowledges the use of OpenAI Codex in the preparation of this work; its role in the calculation and manuscript is described above.
This work is supported in part by the National Natural Science Foundation of China
(NSFC) under Grant No.~1234710148, and in part by the China
Postdoctoral Science Foundation under Grant No.~2023M742098.
\end{acknowledgments}

\appendix
\section{Numerical definitions and checks}

Fourier transforms, the opacity scale, and the background use $\hbar c=0.1973269804~\mathrm{GeV\,fm}$. The overlap routine uses the rounded conversion $0.1973~\mathrm{GeV\,fm}$; replacing it by the full value changes the coherent width by 0.012\%, below the quoted numerical precision.
The dipole opacity uses $A_g=2.373$, $\lambda_g=0.052$, $\mu_0^2=1.428~\mathrm{GeV}^2$, $C=4$, $n_f=4$, and $\Lambda_{\rm QCD}=0.156$ GeV, with initial $xg=A_gx^{-\lambda_g}(1-x)^{5.6}$.
LO gluon-only evolution is integrated by a 40-step RK4 procedure on a 200-point rapidity grid.
Its common evolved table is evaluated separately at each channel's $x$, avoiding a fixed-$x$ opacity reuse.
The baseline table has 28 points from $\mu_0^2$ to $600~\mathrm{GeV}^2$ and linear interpolation in $\mu^2$; above the tabulated range the gluon density is held at its endpoint while the running coupling continues to vary.
Increasing to 42 points and $1200~\mathrm{GeV}^2$ tests that cutoff and interpolation together.
The overlap parameters $(m_q,M_V,N_T,\mathcal R^2)$ are $(1.4,3.097,0.578,2.3)$ for $\Jpsi$ and $(4.2,9.460,0.481,0.57)$ for $\Ups$, with masses in GeV and $\mathcal R^2$ in $\mathrm{GeV}^{-2}$.
The production fit does not use a quark-singlet evolution sector.

The integration grids are $b\in[0.001,15]$ fm (220 points), $r\in[0.001,1.8]$ fm (150), longitudinal nuclear coordinate $z_N\in[-18,18]$ fm (401), and dipole momentum fraction $z\in[0.002,0.998]$ (64).
The raw $|t|$ grid has 200 points through $0.145~\mathrm{GeV}^2$.
The resolution integral over a measured bin uses the difference of two normal cumulative distribution functions, not the spectrum at the bin center.
The relative-coordinate Bessel function is evaluated with its convergent series through order 20; comparison with direct $J_0$ quadrature gives a relative maximum difference below $10^{-14}$ at the reference point.
The overlap exchange check has the same numerical accuracy.

As a coordinate-convention diagnostic, replacing the midpoint phase factor $J_0[(z-1/2)qr]$ by $J_0[(1-z)qr]$ while retaining the midpoint thickness gives a local linear skin shift of $\LegacyPhaseBias$ fm when the resulting spectra are fitted with the midpoint model at the kinematics of Table~\ref{tab:kinematics}.
This comparison deliberately mixes coordinate conventions.
A consistent change to endpoint coordinates would also transform the thickness argument, so the quoted shift is a diagnostic of that mismatch, not an additional physical uncertainty.

The coherent Fisher width is $\CoherentSigma$ fm on the reference grid and $\FineGridSigma$ fm with 330 impact-parameter points, 225 dipole radii, 601 nuclear longitudinal points, 96 momentum-fraction points, and 300 raw-$t$ points.
Halving the reference finite-difference steps gives $\HalfStepSigma$ fm.
The wider evolution table gives $\WideEvolutionSigma$ fm.
The finite-difference steps in the Fisher calculation are 0.006 fm for the skin, $0.005~\mathrm{fm}^2$ for $B_{\rm int}$, 0.02 for $\ln O$, 0.03 for $\gamma$, 0.008 fm for $a_n$ and $R_p$, 0.01 for $\kappag$, 0.04 for $\eta_t$, and 0.02 for either overlap-size nuisance.
The 13-parameter coherent information matrix has full rank; direct inversion is used.
Using 22, 44, and 66 measured bins gives 0.02482, 0.02447, and 0.02438 fm, respectively.
These small changes include the binning dependence of the explicitly per-bin 0.5\% auxiliary uncertainty; they are not evidence that an arbitrary detector covariance is negligible.

\section{Background provenance and interval construction}

The Good--Walker template samples 208 Woods--Saxon nucleon centers with half-density radius 6.62 fm and diffuseness 0.55 fm.
Each nucleon contains three Gaussian hot spots, with $B_{\rm center}=3.0~\mathrm{GeV}^{-2}$ and $B_{\rm hotspot}=0.7~\mathrm{GeV}^{-2}$.
Mean-one log-normal weights have log standard deviation 0.5; each configuration is recentered with these weights.
The coherent and incoherent spectra are the squared mean amplitude and its variance, respectively.
The seed is 7 and the sample size is 4000.
The grid has 100 points in $q\in[0.001,2.0]~\mathrm{fm}^{-1}$, reaching $|t|=0.15575~\mathrm{GeV}^2$, beyond both the fitted interval and the resolution-convolution grid.
No high-$|t|$ log-spectrum extrapolation is used.
The template fixes a background shape independently of the signal geometry; it is not fitted as a self-consistent model of the same nuclear configurations, and template Monte Carlo uncertainty is not separately propagated.
Its normalization and tilt uncertainties are the explicitly chosen surrogates in Eq.~(\ref{eq:background}).

The repeated-experiment seed is 20261027.
Six response and density scenarios are generated sequentially with one random-number stream; the edge scenario is an additional run initialized with the same seed.
For every trial the bin counts and all auxiliary observations are resampled independently.
The reference fit profiles the bin multipliers exactly and solves the remaining direct nonlinear problem with bounded least squares on signed Poisson-deviance residuals plus auxiliary residuals.
Broad numerical bounds include $-0.2<s<0.7$ fm and $-1<\gamma<1$; they are not informative skin priors.
The local expected-information covariance at the fitted physics point and unit bin-response multipliers uses the analytic bin-response Schur complement, $V_{ii}=m_i+(0.005m_i)^2$.
Thus all background derivatives and nuisance correlations enter the reported width.
The per-trial output records the fitted skin, width, coverage indicator, optimizer status, and optimality diagnostic.
The binomial standard error $\sqrt{\hat p(1-\hat p)/200}$ characterizes Monte Carlo precision; a few percentage points of disagreement with 68.27\% alone do not resolve a coverage defect.
Asimov fits diagnose deterministic bias without that sampling error.

The exact edge degeneracy is also checked numerically: the bin counts at nuclear skin 0.17 fm with $\delta_g=0.03$ fm agree with those at skin 0.20 fm with $\delta_g=0$ to machine precision.
The corresponding leading moment changes by $0.13651~\mathrm{fm}^2$, consistent with the linear tolerance in Eq.~(\ref{eq:tolerance}).
This verification concerns the defined optical model; it does not assign an empirical prior to the gluon-edge displacement.

\clearpage
\bibliographystyle{apsrev4-2}

\bibliography{clean_refs}

\end{document}

%% file: data/numbers.tex
\newcommand{\CoherentSigma}{0.02458}
\newcommand{\BackgroundSigma}{0.03120}
\newcommand{\GammaRho}{0.75045}
\newcommand{\OpacityRho}{0.04850}
\newcommand{\LeadingMoment}{20.55191}
\newcommand{\FullLeadingMoment}{20.92191}
\newcommand{\ProcessMoment}{21.90392}

\newcommand{\EdgeThreeSigma}{0.04328}
\newcommand{\EdgeTwoSigma}{0.03706}
\newcommand{\LegacyPhaseBias}{0.00748}
\newcommand{\GammaTight}{0.01768}
\newcommand{\GammaLoose}{0.03484}
\newcommand{\MomentTolerance}{0.13614}
\newcommand{\MomentCoefficient}{0.22036}
\newcommand{\RadialFreeSigma}{0.02458}
\newcommand{\RadialIndependentSigma}{0.02508}
\newcommand{\FineGridSigma}{0.02458}
\newcommand{\WideEvolutionSigma}{0.02459}
\newcommand{\HalfStepSigma}{0.02457}
\newcommand{\MVBias}{-0.01035}
\newcommand{\NominalCoverage}{68.5}

\newcommand{\GammaCorrectCoverage}{67.0}

\newcommand{\GammaOffsetCoverage}{46.5}

\newcommand{\InsideCoverage}{67.5}

\newcommand{\BCGCCoverage}{70.0}

\newcommand{\RadialCoverage}{66.5}

\newcommand{\EdgeCoverage}{38.5}

%% file: data/kinematics.tex
$\Jpsi_T(0)$ & 0.002 & 0.666 & 69.26 & $5\times10^5$ \\
$\Jpsi_T(10)$ & 0.004 & 0.680 & 69.92 & $5\times10^4$ \\
$\Ups_T(0)$ & 0.020 & 0.621 & 66.90 & $1.7\times10^3$ \\

%% file: data/closure_table.tex
Nominal & +0.0000 & 0.0315 & -0.0031 & 68.5 $\pm$ 3.3 & 0.00 \\
$\gamma=-0.26$, centered & +0.0000 & 0.0301 & +0.0031 & 67.0 $\pm$ 3.3 & 0.00 \\
$\gamma=-0.26$, offset & +0.0300 & 0.0306 & +0.0297 & 46.5 $\pm$ 3.5 & 0.80 \\
Inside opacity & +0.0124 & 0.0312 & +0.0074 & 67.5 $\pm$ 3.3 & 0.08 \\
Published bCGC & -0.0045 & 0.0314 & +0.0007 & 70.0 $\pm$ 3.2 & 0.29 \\
$\eta_A=0.05$, omitted & -0.0097 & 0.0320 & -0.0086 & 66.5 $\pm$ 3.3 & 0.86 \\
$\delta_g=0.03$ fm, omitted & +0.0300 & 0.0310 & +0.0335 & 38.5 $\pm$ 3.4 & 0.00 \\